\documentstyle[psfig,conf_iap,12pt]{article}
\def\zform{{${\rm z}_{\rm form}$}}
\begin{document}
\heading{The chemical evolution of galaxies causing \\
damped Ly$\alpha$ absorption} 
\par\medskip\noindent
\author{Ulrich Lindner$^{1}$, Uta Fritze -- v. Alvensleben$^{1}$,
Klaus J. Fricke$^{1}$}
\address{Universit\"atssternwarte, Geismarlandstra\ss e 11,
37083 G\"ottingen, \\ Germany}

\begin{abstract}
We have compiled all available data on chemical abundances in 
damped Lyman alpha absorption systems for comparison with 
results from our combined chemical and spectrophotometric
galaxy evolution models. Preliminary results from 
{\bf chemically consistent} calculations are in  
agreement with observations of damped Ly$\alpha$ systems.
\end{abstract}
\section{Models of galaxy evolution}
Our model is most briefly described as kind of a synthesis 
of a modified Tinsley model for the chemical evolution 
(SN$\,$Ia treated \`a la Matteuchi) and a 
Bruzual--like model for the spectral evolution of the 
stellar population.
Models have been improved to be {\bf chemically consistent}, i.e.
we use evolutionary tracks, 
yields, lifetimes and remnant masses for the metallicity
calculated at the time of birth for respective stars.
Metallicity dependent stellar yields
and remnant masses for massive stars ($M \ge 12*M_\odot$) 
are taken from Weaver \& Woosley\cite{WW}.
For lower mass stars data from van den Hoek \& Groenewegen\cite{vdH} 
are incorporated. Updated SN$\,$Ia contributions are taken from 
Nomoto et al.\cite{Nomoto}.
Adopting any cosmological model ($H_0$, $\Omega_0$, 
$\Lambda_0$) and epoch of galaxy formation (\zform)
we obtain redshift dependent quantities (ISM abundances,
spectra, etc.) which can be compared to observations.

\begin{figure} \vskip -2.25cm
\centerline{\vbox{\psfig{figure=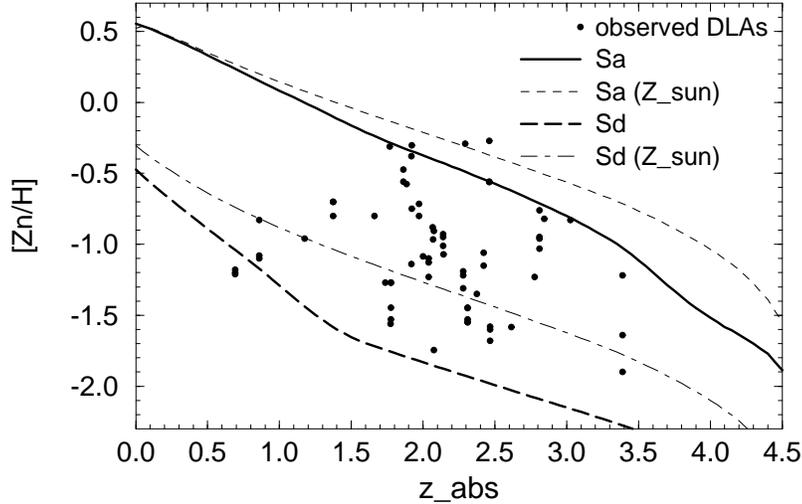,height=9.5cm}}}
\vskip -0.75cm
\caption[]{Comparison of zinc abundances as a function of redshift
for observed DLA systems with chemically consistent calculated models
and models using solar abundances throughout (Z\_sun).}
\end{figure}

\section{Preliminary results and Conclusion}
Figure~1 shows preliminary results for the abundances of zinc.
A considerable part of observed DLA abundances\cite{Pet}\cite{Lu} 
lies between our Sa and Sd models. In particular the chemically consistent
Sd model fits the observational data better than the
model calculated with solar abundances!
The rough agreement of element abundances from our
models with those for high redshift galaxies (DLA systems at
$z > 1$) is an important result, because the SFR we use
has been derived from comparison with photometric data of
low redshift galaxies ($z < 1$).
The appreciable number of precise DLA abundances now available
enables us for the first time to compare results of our 
chemical evolution models with high redshift ($z > 1$) galaxies.

%
\vfill

\begin{iapbib}{99}{
\bibitem{Lu} Lu, L., et al., 1996, ApJS 107, 475
\bibitem{Nomoto} Nomoto, K., et al., 1997, preprint astro-ph/9706025
\bibitem{Pet} Pettini, M., et al., 1997, ApJ 486, 665 
\bibitem{vdH} van den Hoek, L.B., \& Groenewegen, M.A.T., 1997, A\&A Suppl. 123, 305
\bibitem{WW} Woosley, S.E. \& Weaver, T.A., 1995, ApJS 101, 181}\end{iapbib}
\end{document}